\documentclass[twocolumn,aps,prl,floatfix,superscriptaddress]{revtex4}

\usepackage[english]{babel}

\usepackage[letterpaper,top=2cm,bottom=2cm,left=2cm,right=2cm,marginparwidth=1.7cm]{geometry}

\usepackage{amsmath}
\usepackage{graphicx}
\usepackage[colorlinks=true, allcolors=blue]{hyperref}
\usepackage[normalem]{ulem}
\usepackage{comment}

\newcommand{\PbPb}{Pb+Pb}
\newcommand{\AuAu}{Au+Au}

\newcommand{\pp}{\ensuremath{p}+\ensuremath{p}}

\newcommand{\sqrtsNN}{\ensuremath{\sqrt{s_{_\mathrm {NN}}}}}

\newcommand{\IAA}{\ensuremath{I_{\mathrm{AA}}}}
\newcommand{\IAALBT}{\ensuremath{I_{\mathrm{AA}}^{\mathrm{LBT}}}}
\newcommand{\IAAbaseline}{\ensuremath{I_{\mathrm{AA}}^{\mathrm{baseline}}}}

\newcommand{\pT}{\ensuremath{p_\mathrm{T}}}
\newcommand{\pTjet}{\ensuremath{p_\mathrm{T,jet}}}
\newcommand{\pTtrig}{\ensuremath{p_\mathrm{T,trig}}}
\newcommand{\pTpy}{\ensuremath{p_{\mathrm{T}}^{\mathrm{PYTHIA}}}}

\newcommand{\gev}{GeV/$c$}

\begin{document}

\title{Deciphering yield modification of hadron-triggered semi-inclusive recoil jets in heavy-ion collisions}

\author{Yang He}\email[]{yanghe@rcf.rhic.bnl.gov}\affiliation{Institute of Frontier and Interdisciplinary Science, Shandong University, Qingdao, 266237, China}
\author{Maowu Nie}\email[]{maowu.nie@sdu.edu.cn}\affiliation{Institute of Frontier and Interdisciplinary Science, Shandong University, Qingdao, 266237, China}\affiliation{Key Laboratory of Particle Physics and Particle Irradiation, Ministry of Education, Shandong University, Qingdao, Shandong, 266237, China}
\author{Shanshan Cao}\email[]{shanshan.cao@sdu.edu.cn}\affiliation{Institute of Frontier and Interdisciplinary Science, Shandong University, Qingdao, 266237, China}\affiliation{Key Laboratory of Particle Physics and Particle Irradiation, Ministry of Education, Shandong University, Qingdao, Shandong, 266237, China}
\author{Rongrong Ma}\email[]{marr@bnl.gov}\affiliation{Physics Department, Brookhaven National Laboratory, Upton, New York 11973, USA}
\author{Li Yi}\email[]{li.yi@sdu.edu.cn}\affiliation{Institute of Frontier and Interdisciplinary Science, Shandong University, Qingdao, 266237, China}\affiliation{Key Laboratory of Particle Physics and Particle Irradiation, Ministry of Education, Shandong University, Qingdao, Shandong, 266237, China}
\author{Helen Caines}\email[]{helen.caines@yale.edu}\affiliation{Yale University, New Haven, Connecticut 06520, USA}

\begin{abstract}
In relativistic heavy-ion collisions, a hot and dense state of matter, called the Quark-Gluon Plasma (QGP), is produced. Semi-inclusive jets recoiling from trigger hadrons of high transverse momenta ($p_{\mathrm{T}}$) can serve as an effective probe of the QGP properties, as they are expected to experience jet quenching when traversing the QGP. Recent experimental results on the ratio of recoil jet yields normalized by the trigger counts in heavy-ion collisions to that in $p$+$p$ collisions ($I_{\mathrm{AA}}$) pose an unexpected challenge in its interpretation. It is observed that $I_{\mathrm{AA}}$ rises with the jet $p_{\mathrm{T}}$ and possibly exceeds unity at high $p_{\mathrm{T}}$, while traditionally it is expected that jet quenching would lead to $I_{\mathrm{AA}} < 1$. To address this challenge, we utilize the Linear Boltzmann Transport (LBT) model to simulate jet transport in the QGP, and study the effect of jet quenching for high-$p_{\mathrm{T}}$ triggers and recoil jets separately on $I_{\mathrm{AA}}$. We find that the quenching of the colored triggers alone is responsible for the rising trend and larger-than-unity value observed experimentally.
\end{abstract}


\maketitle

\section{Introduction}
\label{introduction}

Energetic partons, produced from hard scatterings with large momentum transfer $Q^{2}$, undergo a sequence of splittings (or parton showers) and form collimated sprays of hadrons known as jets. Since the first observation of jets~\cite{Hanson:1975fe}, they have played a vital role in revealing properties of Quantum Chromodynamics (QCD) both in vacuum and at extremely high temperature and density achieved in relativistic heavy-ion collisions~\cite{Sapeta:2015gee, Larkoski:2017jix,Wang:1992qdg,Qin:2015srf,Majumder:2010qh,Blaizot:2015lma,Cao:2020wlm,Cunqueiro:2021wls,Cao:2022odi}. In the latter case, jet partons lose a portion of their energies while traversing the QCD medium before fragmenting into hadrons. The significant suppression of jet production in nucleus-nucleus (A+A) collisions compared to \pp\ collisions, known as jet quenching, was considered smoking-gun evidence of the formation of the color deconfined Quark-Gluon Plasma (QGP) in heavy-ion collisions~\cite{Shuryak:2008eq,Busza:2018rrf,Harris:2023tti}. Over the past three decades, significant efforts have been devoted to understanding the dynamics of jet-QGP interactions~\cite{Bass:2008rv,JET:2013cls} and inferring QGP properties from jet quenching measurements~\cite{Feal:2019xfl,Liu:2023rfi,Karmakar:2023ity}. 

The large, fluctuating background in heavy-ion collisions makes measurements designed to reveal jet quenching highly challenging. Various strategies have been developed to mitigate these background fluctuation effects, but usually with a price, such as limited kinematic coverage or biases in jet fragmentation~\cite{ALICE:2019qyj, ATLAS:2023hso, CMS:2021vui, STAR:2020xiv}. One particular approach to remove these limitations is to measure jets recoiling from hadron triggers with high transverse momentum (\pT), referred to as semi-inclusive hadron+jet measurements~\cite{ALICE:2015mdb, ALICE:2023jye, STAR:2017hhs, STAR:2023ksv, STAR:2023pal}. The presence of the high-\pT\ trigger can isolate the hard process of interest in heavy-ion collisions, and thus strongly suppresses the background contributions. The semi-inclusive nature of the approach further enables a complete removal of the remaining combinatorial background on an ensemble basis. The combination of the two aspects makes it possible to measure jets with large radii and down to very low \pT\ (3-5 \gev), a unique and critical phase space for a comprehensive understanding of the jet-QGP interactions~\cite{ALICE:2015mdb, ALICE:2023jye, STAR:2017hhs, STAR:2023ksv, STAR:2023pal}. 

To quantify the parton energy loss, the yield of recoil jets per trigger hadron is measured in A+A collisions and compared to that in \pp\ collisions where the QGP is not expected to be formed. The corresponding ratio is denoted: 
\begin{equation} 
    I_{\rm{AA}} = \frac{(1/N_{\rm{trig}})(dN_{\rm{jet}}/d\pTjet)|_{\rm{AA}}}{(1/N_{\rm{trig}})(dN_{\rm{jet}}/d\pTjet)|_{pp}}.
\label{eq:IAA}
\end{equation}
Due to the steeply falling spectrum of the trigger hadrons, it was perceived that they are dominantly produced close to the edge of the QGP and escape it with little energy loss. This is usually referred to as the ``surface bias"~\cite{Muller:2002fa, Renk:2006nd}. Consequently, the recoil jets traverse a longer path through the QGP than the inclusive jet sample, resulting in significant energy loss or a shift in \pTjet. This is expected to manifest as $\IAA < 1$, given that the baseline of no jet quenching is usually considered to be $\IAA = 1$. It is also worth highlighting that Ref. \cite{Milhano:2015mng} concluded that fluctuations in the parton energy loss, rather than the surface bias, dominates the observed di-jet momentum imbalance at the LHC.

A recent measurement with trigger hadrons of $20 < \pTtrig < 50$ \gev\ in \PbPb\ collisions at \sqrtsNN\ = 5.02~TeV by the ALICE experiment at the Large Hadron Collider (LHC) reveals an unexpected behavior of $\IAA > 1$ for charged-particle jets above 110 \gev\ \cite{ALICE:2023jye}. This raises the question whether jet energy loss in the QGP necessarily results in $\IAA < 1$, and how to interpret the \IAA\ measurements in general. While the hadron-triggered jets have been explored in a few theoretical studies~\cite{Qin:2009bk,Chen:2016vem} and the recent experimental observation has been compared to several model calculations~\cite{Zapp:2008gi, JETSCAPE:2021ehl, Casalderrey-Solana:2014bpa}, the origin of the unexpected behavior of $\IAA > 1$ has not been clearly identified. In this work, we will study the surface bias phenomenon in detail and explore how it affects the interpretation of the experimental data on hadron-triggered recoil jets, based on a Linear Boltzmann Transport (LBT) model. 
%This paper is organized as follows: it begins with a concise introduction to the LBT model. Next, we compare the model’s results of energy loss to experimental data obtained from the ALICE collaboration, and compare inclusive jet, single particle $R_{\rm AA}$ and hadron trigger recoil jet $I_{\rm AA}$. Finally, we employ the LBT model to calculate jet suppression with a devised sample, as a means of illustrating how does trigger energy loss affect $I_{AA}$. 

\section{LBT model}
\label{LBT_model}

We use the PYTHIA8 event generator~\cite{Sjostrand:2014zea,Sjostrand:2006za} to simulate the creation of energetic partons from the initial hard scatterings and their subsequent showering in vacuum. The resulting final-state partons from PYTHIA are fed into the LBT model~\cite{Cao:2016gvr,Luo:2023nsi} for further evolution inside the QGP medium. This includes both elastic and inelastic scatterings between jet partons and medium constituent partons, and is known as medium modification of the parton shower. The QGP is described by the relativistic hydrodynamic model VISHNU with an average profile~\cite{Song:2007fn,Song:2007ux,Qiu:2011hf}.
%The LBT model is developed for studying jet evolution inside the QGP, encompassing both elastic and inelastic scatterings between jet partons and medium constituents~\cite{Cao:2016gvr,Luo:2023nsi}. The jet parton production from the initial hard scatterings is simulated using the PYTHIA8 event generator~\cite{Sjostrand:2014zea,Sjostrand:2006za}, while the evolution of the QGP is described by the relativistic hydrodynamic model VISHNU with an average profile~\cite{Song:2007fn,Song:2007ux,Qiu:2011hf}. 

During the QGP phase, the  phase space distribution of jet partons, $f_a(t,\vec{x}_a,\vec{p}_a$), evolves according to the Boltzmann equation as:
\begin{equation} 
	p_a \cdot \partial f_a = E_a\left[C^{\rm el}(f_a) + C^{\rm inel}(f_a\right)],
\end{equation}
where $p_a=(E_a,\vec{p}_a)$ is the four-momentum of the jet parton, and $C^{\rm el}$ and $C^{\rm inel}$ are the collision integrals for elastic and inelastic scatterings respectively. From $C^{\rm el}$, one can extract the elastic scattering rate of a single jet parton as
\begin{align}
\label{eq:rate}
\Gamma_a^\mathrm{el}&(E_a,T)=\sum_{b,(cd)}\frac{\gamma_b}{2E_a}\int \prod_{i=b,c,d}\frac{d^3p_i}{E_i(2\pi)^3} f_b(E_b,T) \nonumber\\
&\times [1\pm f_c(E_c,T)][1\pm f_d(E_d,T)] S_2(\hat{s},\hat{t},\hat{u})\nonumber\\
&\times (2\pi)^4\delta^{(4)}(p_a+p_b-p_c-p_d)|\mathcal{M}_{ab\rightarrow cd}|^2,
\end{align}
where the sum runs over all possible scattering channels $ab\rightarrow cd$, with $b$ being a thermal parton from the medium, $c$ and $d$ being the final states of $a$ and $b$ respectively. In the equation above, $\gamma_b$ is the spin-color degree of freedom for $b$, $f_i$ ($i=b,c,d$) takes the Bose/Fermi distribution for gluons/quarks in the medium rest frame, $|M_{ab\rightarrow cd}|^2$ is the scattering amplitude of a $2\rightarrow2$ process that is proportional to $\alpha_\mathrm{s}^2$ (strong coupling strength). The double-theta function $S_2(\hat{s},\hat{t},\hat{u})=\theta(\hat{s}\ge 2\mu_\mathrm{D}^2)\,\theta(-\hat{s}+\mu^2_\mathrm{D}\le \hat{t} \le -\mu_\mathrm{D}^2)$ is introduced to avoid the divergence of the leading-order matrix element, where $\hat{s}$, $\hat{t}$, $\hat{u}$ are the Mandelstam variables and $\mu_{\rm{D}}^2=4\pi\alpha_\mathrm{s}T^2(N_c + N_f/2)/3$ is the Debye screening mass, with $N_c$ and $N_f$ being the color and flavor numbers. 

The inelastic scattering rate is related to the number of medium-induced gluons per unit time as
\begin{equation}
\label{eq:gluonnumber}
\Gamma_a^\mathrm{inel} (E_a,T,t) = \int dzdk_\perp^2 \frac{1}{1+\delta^{ag}}\frac{dN_g^a}{dz dk_\perp^2 dt},
\end{equation}
where the emitted gluon spectrum ${dN_g^a}/(dz dk_\perp^2 dt)$ is taken from the higher-twist energy loss calculation~\cite{Wang:2001ifa,Zhang:2003wk,Majumder:2009ge}, and $\delta^{ag}$ is applied to avoid double counting in evaluating the $g\rightarrow gg$ rate from its splitting function. The medium effect on this gluon spectrum is absorbed in a jet quenching parameter that further relies on $\alpha_\mathrm{s}$. In the end, $\alpha_\mathrm{s}$ is the sole parameter used in this LBT model. With these differential and integrated rates, one can implement Monte-Carlo (MC) simulations of parton-QGP interactions.

We start the interaction at $t_\mathrm{init}=\max(\tau_0,\tau_\mathrm{form})$, where $\tau_0=0.6$~fm/$c$ is the initial time of the hydrodynamic evolution of the QGP, and $\tau_\mathrm{form}=2Ez(1-z)/k_\perp^2$ is the parton formation time extracted from the PYTHIA simulation~\cite{Zhang:2022ctd}, with $E$ being the energy of the parton's ancestor produced from the initial hard scattering, $z$ being the fractional energy taken from its ancestor, and $k_\perp$ being its transverse momentum with respect to its ancestor. Possible nuclear modification of a highly virtual parton before $\tau_\mathrm{form}$ is beyond the applicability of our transport model, and it can be described by medium-modified parton shower models like Q-PYTHIA~\cite{Armesto:2009fj} and MATTER~\cite{Cao:2017qpx}. During parton-QGP scatterings after $t_\mathrm{init}$, we track not only the final states of the primordial jet partons and their emitted gluons, but also thermal partons scattered out of the medium (``recoil partons") and the associated energy depletion inside the medium (``back-reaction" or ``negative partons"). Recoil partons can further scatter with the QGP in the same way as jet partons do. Such interactions are iterated for each parton until it exits the QGP, \textit{i.e.}, when its local temperature is below 165~MeV. In the end, we cluster the final-state partons into jets of various radii ($R$) using the Fastjet package~\cite{Cacciari:2011ma} with the anti-$k_{\rm{T}}$ algorithm \cite{Cacciari:2008gp}, which has been modified to subtract the momenta of ``negative" partons from those of regular ones~\cite{He:2018xjv}. 

The hadronization process has not been implemented for the results presented in this work. Although this may prevent a quantitative comparison to the experimental data, we do not expect it to affect our qualitative conclusion, since the convoluted effects of energy loss and surface bias are already present at the parton level before they fragment into (trigger) hadrons. With this setup, we will use $\alpha_\mathrm{s}=0.3$, which provides a reasonable description of yield suppression for inclusive jets measured at the Relativistic Heavy Ion Collider (RHIC)~\cite{Zhang:2022ctd}.

\section{Nuclear modification of parton-triggered jets}
\label{Results}

\begin{figure}[tbp!]
	\centering
	\includegraphics[scale=0.45]{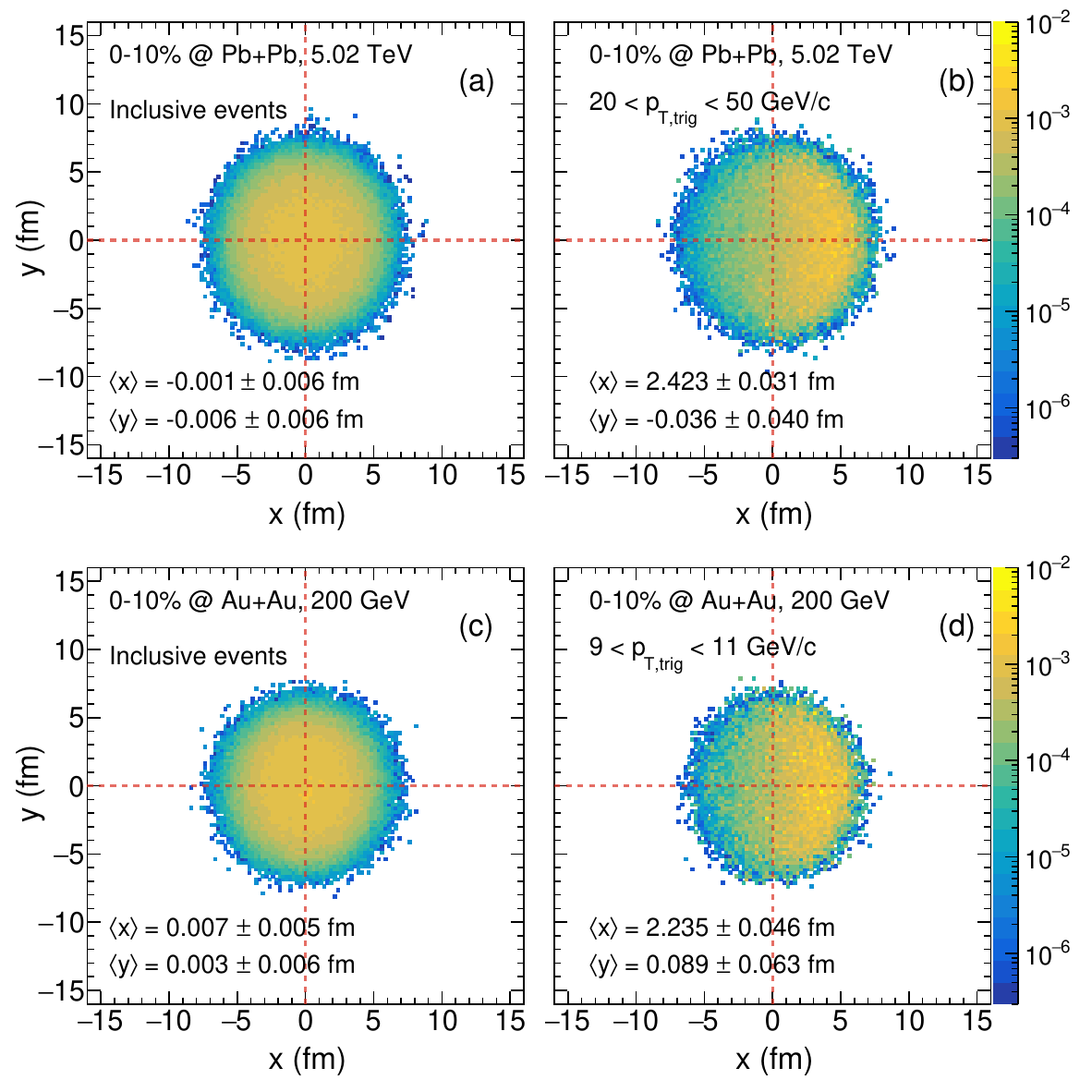}
	\caption{Distributions of parton production vertices in the transverse plane. Top panel: inclusive events (a) and events containing a trigger with $20<\pT<50$~\gev\ (b) in \PbPb\ collisions at \sqrtsNN\ = 5.02 TeV. Bottom panel: inclusive events (c) and events containing a trigger with $9<\pT<11$~\gev\ (d) in \AuAu\ collisions at \sqrtsNN\ = 200~GeV. All distributions are normalized to 1.
    }
	\label{fig1:vertices}
\end{figure}

To study the surface bias of high-\pT\ colored triggers, we start with examining the initial production vertices of these triggers in the transverse ($x$-$y$) plane for 0-10\% (central with small impact parameter) \AuAu\ collisions at \sqrtsNN\ = 200~GeV and \PbPb\ collisions at \sqrtsNN\ = 5.02~TeV in Fig.~\ref{fig1:vertices}. To help visualize the degree of surface bias, we rotate the coordinate vector in the transverse plane ($\vec{r}_\mathrm{T}$) of each initial hard scattering (determined by a MC-Glauber model \cite{Loizides:2017ack}) such that the transverse momentum of the final-state trigger aligns with the $+\hat{x}$ direction, defined by the impact parameter between the two colliding nuclei. As a reference, the distributions of the inclusive parton production vertices are also shown in panels (a) and (c), which are symmetric against the center $(x,y)=(0,0)$ as expected. For triggering partons, {\it i.e.}, those within a specific high-\pT\ range after traversing the QGP, their production vertices are clearly skewed away from the center, as shown in panels (b) and (d). The deviation from the center is about 2 fm, inline with the perceived surface bias. However, there are still a large fraction of triggers that lose energy before emerging from the medium, which have a crucial impact on the interpretation of the \IAA\ measurement, as discussed later. 

\begin{figure}[tbp!]
	\centering
	\includegraphics[scale=0.5]{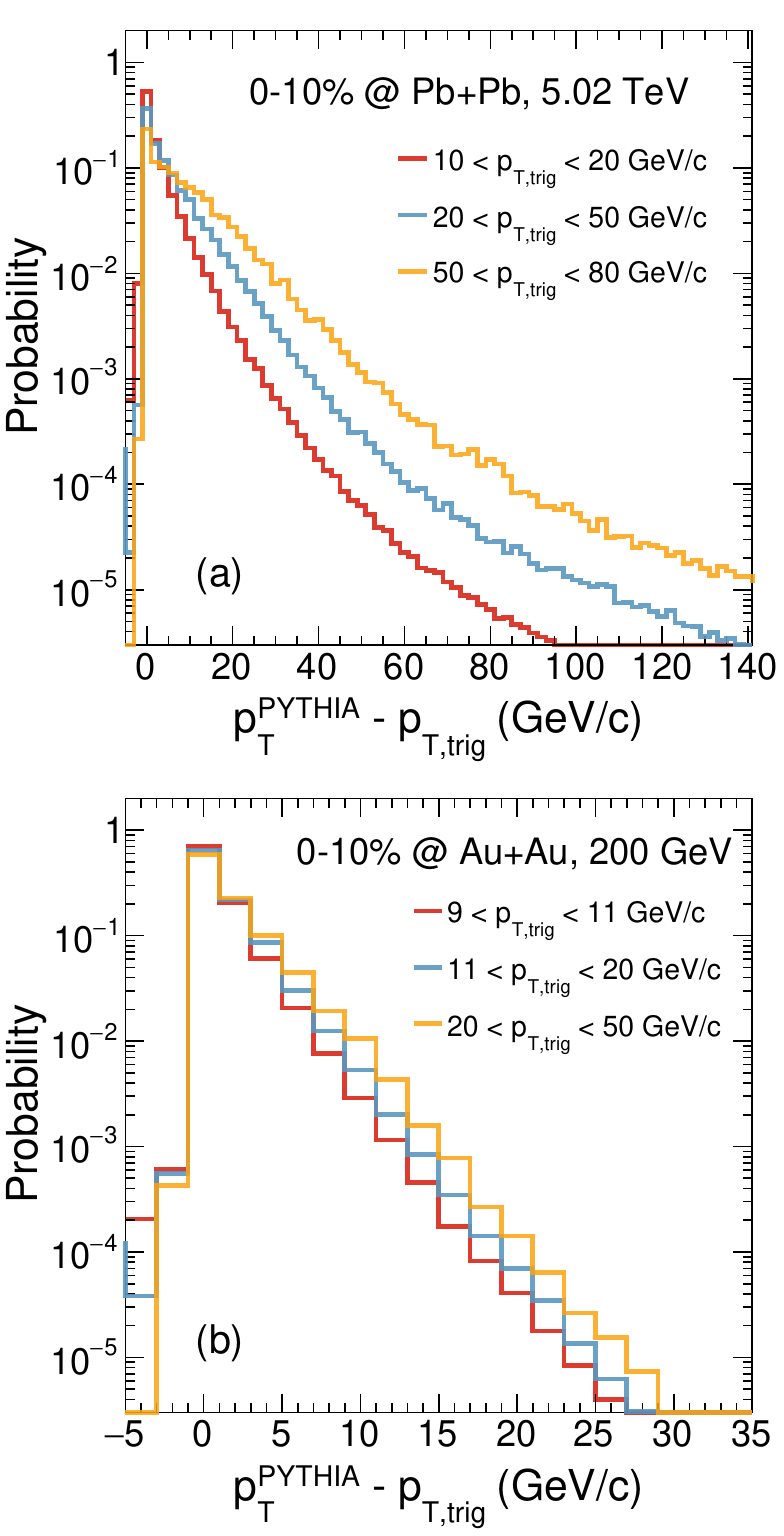}
 	\caption{Distributions of the energy loss for parton triggers inside the QGP within various \pTtrig\ intervals in 0-10\% \PbPb\ (a) and \AuAu\ (b) collisions.
    }
	\label{fig2:energyloss}
\end{figure}

To further quantify the energy loss of trigger partons, distributions of their lost energies are shown in Fig. \ref{fig2:energyloss} for various \pTtrig\ ranges. 
Here, for each trigger parton appearing in an LBT event, we search for its origin from the corresponding PYTHIA event. The origin is identified as the hardest parton in the PYTHIA event within a cone of 0.2 with respect to the trigger parton from LBT. The energy loss is then defined as the $\pT$ difference between a given trigger parton in LBT and its origin: $\Delta \pT\ =\pTpy\ - \pTtrig$. Since there could be mismatch between the trigger parton and its origin due to a large angle scattering, a negative $\Delta \pT$ is occasionally observed but with a very small probability. On the other hand, rare, but extremely large, energy losses could originate from medium-induced hard splittings. It is also worth pointing out that energy losses could become greater than the trigger partons' energies after escaping the QGP, {\it i.e.,} $\Delta \pT > \pTtrig$, as a result of the original partons interacting significantly with the medium. From Fig.~\ref{fig2:energyloss}, we note only about 37\% of the triggers lose little energy (less than 1~\gev) for the selection of $20<\pTtrig<50$~\gev\ in central \PbPb\ collisions, while this fraction is about 70\% for $9<\pTtrig<11$~\gev\ in central \AuAu\ collisions. This is inline with the observation in Fig.~\ref{fig1:vertices} that a large fraction of triggers is produced away from the edge of the medium and thus can lose a significant amount of energy. Compared to the 20-50 \gev\ triggers in 5.02 TeV \PbPb\ collisions, the deviation from the center is smaller for 9-11 \gev\ triggers in 200 GeV \AuAu\ collisions (Fig. \ref{fig1:vertices} (b) and (d)) despite the larger fraction that lose no energy. This is due to a hotter and denser QGP being created at 5.02 TeV which leads to larger energy loss per unit pathlength. In Fig.~\ref{fig2:energyloss}, we also compare different \pTtrig\ ranges. Contrary to naive expectations of a stronger surface bias or less energy loss for higher-\pT\ triggers, the opposite is observed. This is understood as being due to the harder (flatter) parton spectrum at higher \pT. For this reason, partons within a higher \pTtrig\ range receive larger contributions from even higher-\pT\ partons with stronger energy loss, and thus experience weaker surface bias. 

\begin{figure*}[htp!]
	\centering
	\includegraphics[width=0.96\textwidth]{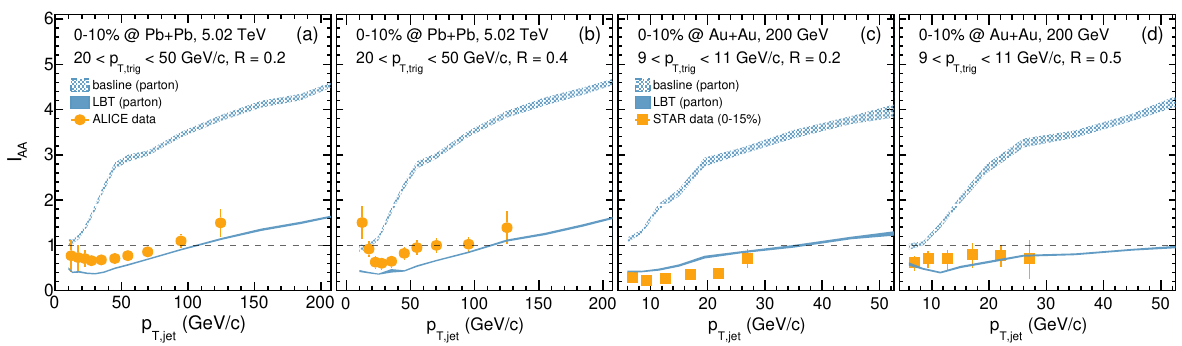}
	\caption{Left two panels: the parton-triggered partonic jet \IAA\ from the LBT calculation (solid band) compared to the ALICE data on charged hadron-triggered charged-particle jets in 0-10\% \PbPb\ collisions at \sqrtsNN\ = 5.02~TeV for jet radii of $R$ = 0.2 (a) and 0.4 (b). The \IAAbaseline\ (hatched bands) is also shown for comparison. Right two panels: similar to the left two panels, but for \IAA\ in 0-10\% Au+Au collisions at \sqrtsNN\ = 200~GeV compared to STAR results of $\pi^{0}$-triggered charged-particle jets in 0-15\% centrality interval for jets with $R$ = 0.2 (c) and 0.5 (d). The bands represent statistical errors in model calculations.}
	\label{fig3:triggereffect}
\end{figure*}

With the improved knowledge on surface bias for trigger partons, we will further explore its effects on the recoil jets next. In Fig.~\ref{fig3:triggereffect}, we compare our model calculation on the nuclear modification of parton-triggered partonic jets to the measurements of charged hadron-triggered charged-particle jets in \PbPb\ collisions at \sqrtsNN\ = 5.02~TeV~\cite{ALICE:2023jye} and $\pi^{0}$-triggered charged-particle jets in \AuAu\ collisions at \sqrtsNN\ = 200~GeV~\cite{STAR:2023ksv, STAR:2023pal}. Here, \IAALBT\ is defined as the ratio of the recoil jet spectra per trigger between LBT (with energy loss) and PYTHIA (without energy loss). Recoil jets are selected within the azimuthal region of $(3\pi/4,5\pi/4)$ for \AuAu\ calculation and $(\pi-0.6, \pi+0.6)$ for \PbPb\ with respect to the trigger particle, in order to match those used experimentally \cite{ALICE:2023jye,STAR:2023ksv, STAR:2023pal}. As seen in Fig.~\ref{fig3:triggereffect}, the rising trend of \IAA\ with \pTjet, especially as observed in the ALICE data, can be qualitatively reproduced by the LBT model that takes into account the energy loss of both trigger particles and recoil jets. \IAALBT\ is below one at low \pTjet, an expected signal of jet quenching, but increases to be above one at high \pTjet. While the increase of \IAA\ with \pTjet\ could be partly understood from the flatter spectra at higher \pTjet, $\IAA > 1$ at high-\pTjet\ is not expected from the general picture of jet quenching. 

To understand this puzzling observation from both experimental data and our LBT calculations, we construct a hybrid sample, {\it i.e.}, for each trigger parton in LBT, recoil jets are constructed from the corresponding PYTHIA event within the same aforementioned azimuthal angle region relative to the trigger. Using this hybrid sample, we define the ``true" baseline of no jet quenching ($\IAA^\mathrm{baseline}$) according to Eq.~(\ref{eq:IAA}), where the numerator is constructed with $N_\mathrm{trig}^\mathrm{LBT}$ that includes the energy loss effect and recoil jets in PYTHIA that do not. As shown in Fig.~\ref{fig3:triggereffect}, $\IAA^\mathrm{baseline}$ is close to one at low \pT\ but rises above one as \pTjet\ increases. Since the same \pTtrig\ range is used for measuring recoil jet spectra in A+A and \pp\ collisions, the energy loss of trigger particles in A+A collisions results in them having originated from higher-$Q^2$ processes, and thus a harder recoil jet spectrum than that in \pp\ collisions. This leads to a significant increase in $\IAA^\mathrm{baseline}$ with \pTjet, far surpassing one at high \pTjet. Because of the harder spectrum, $\IAAbaseline < 1$ could also occur at very low \pTjet. By comparing \IAALBT\ to \IAAbaseline, a clear suppression of the semi-inclusive recoil jet spectrum in A+A {\it vs.} \pp\ collisions is seen, as originally expected from jet quenching. Therefore, the puzzling increase of \IAA\ observed experimentally can be explained by the energy loss of high-\pT\ colored triggers, {\it i.e.}, $\IAA > 1$ can still signal jet quenching given that the baseline of unquenched jets is no longer at unity as naively expected.

\begin{figure}[tbp!]
	\centering
	\includegraphics[scale=0.35]{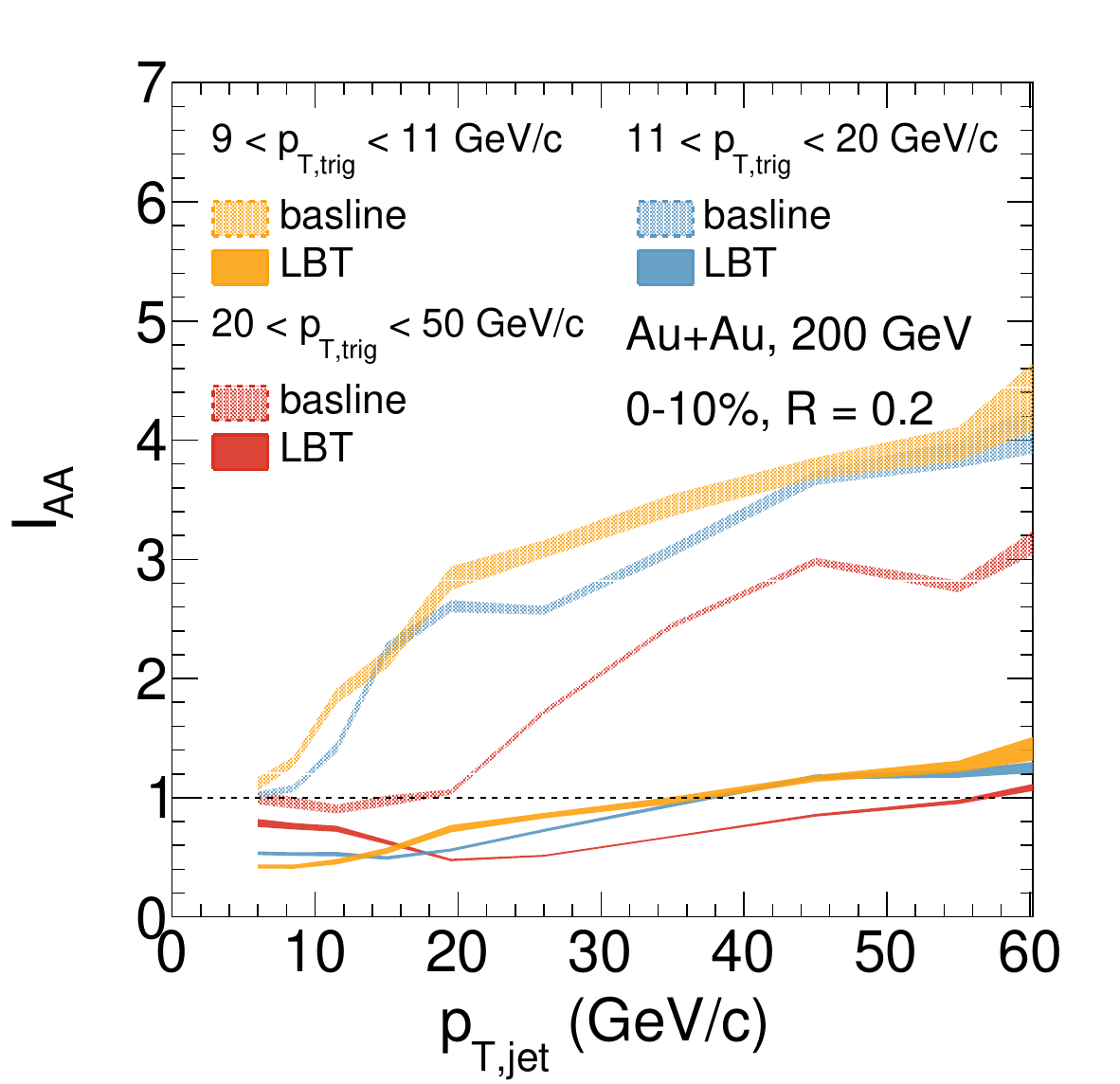}
	\caption{Comparison of \IAAbaseline\ (hatched) and \IAALBT\ (solid) for different trigger \pT\ ranges for $R = 0.2$ jets in 0-10\% \AuAu\ collisions at \sqrtsNN\ = 200 GeV. }
	\label{fig4:triggerbins}
\end{figure}

For a more differential study on jet quenching, we present in Fig. \ref{fig4:triggerbins} the  \IAAbaseline\ and \IAALBT\ of $R=0.2$ jets within three different \pTtrig\ ranges in central \AuAu\ collisions. For all the \pTtrig\ ranges used here, \IAALBT\ is less than one at low \pTjet, and increases above one at high \pTjet. For \IAAbaseline, it stays close to one in the \pTjet\ regime compatible to the \pTtrig\ range before rising quickly up to about 4 at high \pTjet. In the \pTjet\ range examined here, the baseline actually gets closer to one with increasing \pTtrig, which is likely caused by the combined effects of flatter recoil jet spectrum and smaller relative increase in the initial $Q^{2}$ from PYTHIA to LBT events at higher \pTtrig.

\section{Summary}
\label{Summary}
While it was widely assumed that jet quenching should lead to a suppression of hadron-triggered jets at high \pTjet\ ($\IAA < 1$), the recent ALICE data on \IAA\ in central \PbPb\ collisions at \sqrtsNN\ = 5.02~TeV present a challenge in understanding its rising trend and value above one. Using a transport model with parton-level information, we critically re-examine the surface bias and energy loss effects on trigger particles. Because the experimental data is presented at the particle level, direct quantitative comparison to our parton-level results is not possible. Nevertheless, we can draw qualitative conclusions since the effects of energy loss and surface bias already occur at the parton level. We found that while surface bias does exist, there are still a large fraction of triggers being produced deep inside the QGP and suffering from considerable energy loss before being observed. Given that \IAA\ represents the ratio of recoil jet spectra in A+A and \pp\ collisions within the same \pTtrig\ range, the energy loss of triggers enhances \IAA\ to be significantly larger than the initially assumed value of one for unquenched recoil jets. This enhanced baseline indicates that experimental measurements of $\IAA > 1$ could still signal jet quenching. These findings illustrate the complexities in interpreting semi-inclusive hadron+jet measurements since the true baseline accounting for the trigger energy loss could not be obtained experimentally. However, this in turn also provides stronger constraints on model calculations since they need to correctly and coherently describe both the trigger and recoil jet energy loss. Alternatively, one can seek  to use triggers that do not experience jet quenching in the QGP, {\it e.g.} photons, such that the $\IAA$ baseline stays at one, while still maintaining the technical advantages brought by the semi-inclusive approach.

\section*{Acknowledgments}
We thank Nihar R. Sahoo for providing STAR data points, and Peter Jacobs and Saehanseul Oh for insightful discussions. This work was supported by the National Natural Science Foundation of China (NSFC) under Grant Nos.~11890710 (YH, LY), 11890713 (YH, LY), 12105156 (MWN), 12175122 (SC), 14-547 (YH, LY), 2021-867 (SC), National Key R\&D Program of China under Grant No. 2022YFA1604900 (MWN) and Shandong Provincial Natural Science Foundation project ZR2021QA084 (MWN) and by the U.S. DOE Office of Science under contract Nos. DE-SC0012704 (RM), DE-FG02-10ER41666 (RM), DE-AC02-98CH10886 (RM) and DE-SC004168 (HC).

%\bibliographystyle{elsarticle-num}
%\bibliographystyle{h-physrev5}
%\bibliographystyle{model1-num-names}
%\bibliography{references}
\bibliography{model_hjet.bbl}

\end{document}